\documentclass[final,english,twocolumn,amssymb,a4paper,nobibnotes,aps,prb,longbibliography]{revtex4-1}
\usepackage{graphicx}

\usepackage{natbib}
\usepackage{amsmath}
\usepackage{amssymb}
\usepackage{appendix}
\usepackage{comment}
\usepackage[dvipsnames]{xcolor}
\usepackage[section]{placeins}
\usepackage{layouts}
\usepackage{bbold}
\definecolor{darkblue}{rgb}{0,0,.65}
\definecolor{darkgreen}{rgb}{0.28,0.41,0.19}

\usepackage[%
    pdfauthor={David J. Luitz},%
  pdfstartview=FitH,%
  breaklinks=true,%
  bookmarks=true,%
  colorlinks=true,%
  anchorcolor=black,%
  citecolor=blue,
  filecolor=black,%
  menucolor=black,%
  urlcolor=darkblue,%
  linkcolor=blue,%
 ]{hyperref}
\usepackage[all]{hypcap} 
\usepackage{braket}

\newcommand{\tr}{\mathrm{Tr}}

\newcommand{\E}{\mathrm{e}}
\newcommand{\I}{\mathrm{i}}

\begin{document}

\title{Lieb Robinson bounds and out of time order correlators in a long range spin chain}%

\author{Luis Colmenarez}
\author{David J. Luitz}
\affiliation{Max Planck Institute for the Physics of Complex Systems, Noethnitzer Str. 38, Dresden, Germany}
\email{dluitz@pks.mpg.de}
\email{lcolmena@pks.mpg.de}
\date{\today}%

\begin{abstract}
Lieb Robinson bounds quantify the maximal speed of information spreading in nonrelativistic quantum systems. We discuss the relation of Lieb Robinson bounds to out of time order correlators, which correspond to different norms of commutators $C(r,t) = [A_i(t),B_{i+r}]$ of local operators.
Using an exact Krylov space time evolution technique, we calculate these two different norms of such commutators for the spin 1/2 Heisenberg chain with interactions decaying as a power law $1/r^\alpha$ with distance $r$. Our numerical analysis shows that both norms (operator norm and normalized Frobenius norm) exhibit the same asymptotic behavior, namely a linear growth in time at short times and a power law decay in space at long distance, leading asymptotically to power law light cones for $\alpha<1$ and to linear light cones for $\alpha>1$. The asymptotic form of the tails of $C(r,t)\propto t/r^\alpha$ is described by short time perturbation theory which is valid at short times and long distances.
\end{abstract}

\maketitle

\section{Introduction}

One of the most general concepts to study dynamical properties of quantum many-body systems is the dynamics of quantum information, generalizing the spreading of all possible types of correlations in the system. 
Of particular interest is the dynamical spreading of local operators \cite{nahum_operator_2018,khemani_operator_2018,von_keyserlingk_operator_2018}, which contains information about all correlation functions composed of these operators. While in relativistic systems, the spreading of information is limited by the speed of light, there is no such strict limit in nonrelativistic quantum mechanics.
However, it was shown by Lieb and Robinson \cite{lieb_finite_1972} that quantum systems with short range interactions exhibit a similar, non-universal speed limit, implying a causal structure. This emergent ``light cone" has important consequences for the behavior of many-body interacting systems such as the area law of entanglement \cite{hastings_area_2007,eisert_colloquium_2010}, the decay of correlations \cite{hastings_decay_2004} and stability of topological order \cite{bravyi_lieb-robinson_2006}, but also for the timescales of thermalization\cite{kaufman_quantum_2016,kastner_n-scaling_2017,luitz_information_2017}.
Recently, a lot of progress has been made in establishing similar speed limits for the spreading of information in systems with long range interactions and it is clear that also in such systems, information can not spread infinitely fast \cite{luitz_emergent_2019,hastings_spectral_2006,foss-feig_nearly_2015,gong_persistence_2014,guo_signaling_2019,else_improved_2018,tran_locality_2019,hauke_spread_2013,santos_cooperative_2016,chen_finite_2019,kloss_spin_2019,chen_quantum_2019-1}, however it is not always clear whether so far established analytical bounds on information spreading are tight for experimentally relevant lattice models.
Power law decaying interactions are present in several quantum simulator platforms such as trapped ions \cite{kim_quantum_2011,blatt_simulations_2012}, Rydberg atoms \cite{saffman_quantum_2010}, ultra cold atoms \cite{bloch_simulations_2012} and superconducting qubits \cite{otten_universal_2016} and it is therefore important to obtain tight bounds on information scrambling and thermalization timescales. 

A useful measure to quantify the spreading of an initially local operator $\hat{ A}_i(t)$ is the commutator with another local operator $\hat{B}_{i+r}$
\begin{eqnarray}
\label{eq:conmutator_def}
\hat C(r,t) = ||[\hat{A}_i(t),\hat{B}_{i+r}]||
\end{eqnarray}
where $\hat{B}_{i+r}$ serves as a probe and the operators $\hat{A}_i$ and $\hat{B}_{i+r}$ act only on sites $i$ and $i+r$ of the system respectively. $\hat{A}_i(t) = \exp{(\I \hat H t)} \hat{A}_i \exp{(-\I \hat H t)}$ is the operator $\hat{A}_i$ under time evolution in the Heisenberg picture and $||.||$ denotes any matrix norm. 
Vanishing $C(r,t)$ indicates that no information has traveled from site $i$ to $i+r$ at time $t$. It should be noted here that this generally depends very little on the choice of the operator in chaotic quantum many-body systems and only fine tuned situations exist, where a dependence on the operator at short times can be observed \cite{luitz_emergent_2019}.

In systems with short range interactions, $C(r,t)$ is bounded within a ``light-cone'' region $t > r/v$ where $v$ is a velocity that depends on the microscopic model. This bound does not represent a strict cutoff, since exponential tails exist outside the light cone \cite{lieb_finite_1972,khemani_velocity-dependent_2018,wang_tightening_2019}.
	
Similarly, analytical bounds have been derived in systems with long range interactions decaying as a power law $1/r^{\alpha}$ with distance $r$. Hastings and Koma suggested a logarithmic bound $t\sim \log(r)$ \cite{hastings_spectral_2006} for any $\alpha$. 

In the case of strongly long ranged systems with small $\alpha<D$ ($D$ is the spatial dimension), the logarithmic bound is dominant  \cite{storch_interplay_2015, guo_signaling_2019}. 
Polynomial light cones have been proposed \cite{foss-feig_nearly_2015,matsuta_improving_2017,else_improved_2018} of the form $t\sim r^{(\alpha-2D)/(\alpha-D+1)}$ in the regime $\alpha > 2D$ which consistently recovers the linear light cone in the short range limit $\alpha\rightarrow\infty$. This bound was tightened to $t\sim r^{(\alpha-2D)/(\alpha-D)}$ \cite{tran_locality_2019}.  
In general $\alpha$ large but finite is consistently found to exhibit asymptotically short-ranged behavior \cite{hauke_spread_2013,luitz_emergent_2019,chen_finite_2019,chen_quantum_2019}.
It was argued by Gong et al. that a linear light cone structure persists for $\alpha>D$ \cite{gong_persistence_2014}, which is also supported by numerical simulations \cite{luitz_emergent_2019}. A stochastic model of operator spreading in long range interacting systems points to linear light cones for $\alpha\geq D+\frac 1 2$ \cite{zhou_operator_2020-1}.
For general quantum state transfer protocols, only a weaker bound for a linear light cone for $\alpha>2D+1$ is valid \cite{kuwahara_strictly_2019,tran_hierarchy_2020}.

The analysis of analytical bounds \cite{lieb_finite_1972,hastings_spectral_2006,foss-feig_nearly_2015,matsuta_improving_2017,gong_persistence_2014,guo_signaling_2019,storch_interplay_2015,tran_locality_2019,kuwahara_strictly_2019,tran_locality_2019-1} is mostly concerned with the  operator norm $||\hat{C}(r,t)||_2$ (the largest singular value of the commutator matrix $\hat{C}(r,t)$ in Eq. \eqref{eq:conmutator_def}) because it encodes the ``worst case'' scenario, namely the fastest spreading modes in the system. On the other hand, numerical simulations of $C(r,t)$ usually employ the square of the normalized Frobenius norm $\|\hat{C}(r,t)\|_F^2$ \cite{luitz_information_2017,chen_quantum_2019,luitz_emergent_2019}, which is the average over the square of its singular values and is directly related to the out of time order correlator (OTOC) \cite{larkin_quasiclassical_1969,maldacena_bound_2016,roberts_lieb-robinson_2016,shenker_black_2014} as shown in Eq. \eqref{eq:otoc_spin}. %
In Ref. \onlinecite{tran_hierarchy_2020} a bound on the Frobenius norm was established and found to be different from the bound on the operator norm. The Frobenius norm, associated with typical states, was shown to exhibit linear light cones for $\alpha>5/2$, while for the operator norm a weaker bound of $\alpha>3$ was found.

In a previous numerical study of long range interacting spin chains\cite{luitz_emergent_2019}, the asymptotic shape of the light cone of the OTOC (normalized Frobenius norm) was considered. In the present work, we are interested instead in the behavior of the operator norm of the commutator to compare the spreading of the \emph{fastest mode} to that of \emph{typical modes} (Frobenius norm). 
Interestingly, our analysis suggests, that both the average and the largest singular value of $\hat{C}(r,t)$ have the same asymptotic behavior: We find a linear growth in time of $\|\hat{C}(r,t)\|_2$ at short times, and a power law decay with distance at long distances with the exponent $\alpha$, which can be understood from perturbation theory.
  
\section{Model and Method}
We study the isotropic one-dimensional Heisenberg XXX model with long range interactions:
\begin{eqnarray}
\label{eq:model}
H = \sum_{i<j}\!\dfrac{J}{|i-j|^{\alpha}}\!\left(\hat{S}^{x}_{i}\hat{S}^{x}_{j}\!+\hat{S}^{y}_{i}\hat{S}^{y}_{j}+\hat{S}^{z}_{i}\hat{S}^{z}_{j}\right),
\end{eqnarray}
where $S^{\gamma}_i=\sigma^{\gamma}_i/2$ are spin $1/2$ operators acting on site $i$, with $\gamma=x,y,z$ ($\sigma^\gamma_{i}$ are the corresponding Pauli matrices). 
The interaction exponent $\alpha$ controls the range of the interactions and we use $J=1$ throughout this paper. We do not use a rescaling of the coupling constant with system size $L$ to make the energy extensive for small $\alpha$, since this essentially only rescales our units of time.
The model \eqref{eq:model} conserves the total magnetization $S_\text{tot}^z=\sum_i \hat{S_i^z}$ and we focus on the largest magnetization sector $S_\text{tot}^z=0$ for even $L$ and $S_\text{tot}^z=\frac 1 2$ for odd $L$, to maximize the accessible system sizes.
The limit $\alpha\rightarrow 0$ corresponds to all-to-all interactions and $\alpha\rightarrow \infty$ is the nearest neighbor limit, which are both integrable points of the model. 
There is also a special integrable point at $\alpha=2$, the so called Haldane-Shastry model \cite{haldane_exact_1988,shastry_exact_1988}. 

For concreteness, we consider the dynamical spreading of the local $\hat S_i^z(t)$ operator, probed by the commutators
\begin{equation}
\label{eq:commutator_sz}
\hat{C}(r,t) = \left[ \hat S_i^z(t), \hat S_{i+r}^z \right].
\end{equation}
We note that due to the SU(2) symmetry of the model, all $S_i^x$, $S_i^y$, $S_i^z$ operators spread in the same way.

\subsection{Matrix norms of the commutator}

In order to quantify the growth of the commutator norm $C(r,t)$, we use \emph{two different matrix norms}. The (normalized) Frobenius norm is defined as 
\begin{equation}
\label{eq:frobeniusnorm}
\|\hat{C}(r,t)\|_F := \sqrt{ \frac{ \tr{\left(\hat{C}(r,t)^\dagger \hat{C}(r,t) \right)}}{\mathcal N} } = \sqrt{ \frac{\sum_i s_i^2}{\mathcal{N} } },
\end{equation}
where $s_i$ are the \emph{singular values} of the commutator $\hat{C}(r,t)$ (and consequently $s_i^2$ the eigenvalues of $\hat{C}(r,t)^\dagger \hat{C}(r,t)$).

The \emph{operator norm} $\|\hat C(r,t)\|_2$, or 2-norm is defined by the largest singular value
\begin{equation}
\label{eq:opnorm}
\|\hat{C}(r,t)\|_2 = \sup_{\psi \in \mathcal{H}} \frac{ \| \hat{C}(r,t) \ket{\psi} \|_2 }{\| \ket{\psi}\|_2} = \max_i s_i.
\end{equation}

Therefore, the normalized Frobenius norm is always smaller than (or equal to) the operator norm
\begin{equation}
\label{eq:norm_ineq}
C_2(r,t) \geq C_F(r,t).
\end{equation}
Where for simplicity we have denoted $C_{2}(r,t)=\|\hat{C}(r,t)\|_{2}$ and $C_{F}(r,t)=\|\hat{C}(r,t)\|_{F}$.

\subsection{Out of time order correlator and relation to Frobenius norm of the commutator}

Expanding the definition of the normalized Frobenius norm \eqref{eq:frobeniusnorm} for the commutator $\hat{C}(r,t) = [\hat S_i^z(t), \hat S_{i+r}^z]$ yields
\begin{eqnarray}
\label{eq:otoc_spin}
C_{F}(r,t)^2 & = &||[\hat{S}^z_i(t),\hat{S}^z_{i+r}]||_F^2 \nonumber \\
	     & = &\frac{1}{8} - \frac{2}{\mathcal{N}} \tr\left(\hat{S}^z_i(t)\hat{S}^z_{i+r}\hat{S}^z_i(t)\hat{S}^z_{i+r}\right).
\end{eqnarray}

The correlation function $\frac{1}{\mathcal{N}}\tr\left(\hat{S}^z_i(t)\hat{S}^z_{i+r}\hat{S}^z_i(t)\hat{S}^z_{i+r}\right)$ is known as the \emph{out of time order correlator} (OTOC) and can be viewed as an infinite temperature four point function, where the partition function is given by the dimension of the Hilbert space $Z=\mathcal{N}$.

In order to study the long distance behavior of this quantity, it is crucial to access large enough system sizes to ensure the convergence of our results in the thermodynamic limit and we therefore use 
dynamical typicality \cite{bartsch_dynamical_2009} for computing the trace which appears in the Frobenius norm $\tr(C(r,t)^\dagger C(r,t))$. This method consists of replacing the trace operation by the expectation value $\langle \psi |\hat{C}(r,t)^\dagger\hat{C}(r,t)|\psi \rangle $ where $\psi$ is a random vector in the Hilbert space drawn from the Haar measure \cite{haar_massbegriff_1933}, and averaging over random vectors $\ket{\psi}$. Eq. \eqref{eq:otoc_spin} is then boiled down to $\tr\left( \hat C(r,t)^\dagger \hat C(r,t) \right) = \langle \psi^{\prime}|\psi^{\prime}\rangle$ , where $|\psi^{\prime}\rangle= C(r,t)|\psi\rangle$, up to an error exponentially small in the system size $L$, requiring a very small number of random vectors (typically $1\dots100$) for large enough systems. The operation $C(r,t)|\psi\rangle$ is performed as a sequence of matrix-vector multiplications and several time propagations $\E^{-\I Ht}|\psi\rangle$ of (intermediate) wave functions $|\psi\rangle$. These propagations can be performed efficiently using massively parallel sparse matrix Krylov space techniques.
Technical details of this method for the calculation of the OTOC are discussed in Refs. \onlinecite{luitz_emergent_2019,luitz_information_2017,hemeryMatrixProductStates2019}.

\subsection{Operator norm of the commutator}

In the present paper, our main focus is on the operator norm (2-norm) of the commutator
\begin{eqnarray}
\label{eq:op_norm}
C_2(r,t) & = &\|[\hat{S}^z_i(t),\hat{S}^z_{i+r}]\|_2,
\end{eqnarray}
which corresponds to the largest eigenvalue (equivalent to the largest singular value of $C(r,t)$) of the Hermitian form of the commutator $\I C(r,t) = \mathrm{i} [\hat{S}^z_i(t),\hat{S}^z_{i+r}]$. 
We use a matrix free implementation of the matrix vector product $\ket{\tilde{\psi}} \leftarrow \I C(r,t)|\psi\rangle$, such that we never have to deal with dense matrices and use the Lanczos algorithm to obtain the largest eigenvalue of $\I C(r,t)$. 

This means that we calculate 
\begin{equation}
\begin{split}
\I C(r,t) \ket{\psi} &= \I \left[\hat S_i^z(t), \hat S_{i+r}^z\right] \ket{\psi} \\
&= \I \hat S_i^z(t) \hat S_{i+r}^z \ket{\psi} - \I  \hat S_{i+r}^z \hat S_i^z(t) \ket{\psi} \\
&= \I \E^{\I \hat H t} \hat S_i^z \ket{\psi_2(t)} - \I \hat S_{i+r}^z \E^{\I \hat H t} \hat S_i^z \ket{\psi(t)} \\
&= \I \E^{\I \hat H t} \ket{\psi_3} - \I \hat S_{i+r}^z \E^{\I \hat H t} \ket{\psi_4} \\
&= \I \ket{\psi_3(-t)} - \I \hat S_{i+r}^z \ket{\psi_4(-t)} \\
&= \I \ket{\psi_3(-t)} - \I \ket{\psi_5} \\ 
& \to \ket{\tilde{\psi}}.
\end{split}
\end{equation}
Here, we have used the replacements $\ket{\psi_2} = \hat S_{i+r}^z \ket{\psi}$, 
$\ket{\psi_3} = \hat S_{i}^z \ket{\psi_2(t)}$,
$\ket{\psi_4} = \hat S_{i}^z \ket{\psi(t)}$,
$\ket{\psi_5} = \hat S_{i+r}^z \ket{\psi_4(-t)}$. 
The matrix-free matrix-vector product involves again forward $\ket{\psi(t)}=\E^{-\I \hat H t} \ket{\psi}$ and backward $\ket{\psi(-t)}=\E^{\I \hat H t} \ket{\psi}$ real time evolution of the wavefunction, very similarly to the case of the OTOC \cite{luitz_information_2017,luitz_emergent_2019}, for which we employ a Krylov space technique for the matrix exponential \cite{nauts_new_1983,park_unitary_1986,luitz_ergodic_2017}. Matrix vector multiplications of $\hat S^z_i$ operators with wave functions are trivial, since these operators are diagonal in the computational $S^z$ basis, and the entire algorithm thus requires only storage of a few vectors.
This method gives access to the largest eigenvalue of the commutator with controlled accuracy up to system size $L=22$ ($\mathcal{N}=705432$ in the zero magnetization sector). We note that for larger $\alpha$ and short distances $r$, the convergence of the Lanczos algorithm is particularly challenging due to small gaps in the spectrum.
Lastly, for treating small systems $L<18$, the calculations were performed using full exact diagonalization.

Throughout this paper, we fix the position of the spreading operator to $i=3$ (the left most is indexed $i=0$) in such a way that distances $r=0,1,\dots,L-4$ are accessible (using open boundaries) and the reflection of the left information front does not interfere with propagation  of the right one (which is the one we study in detail).  

\section{Results}

\begin{figure*}[ht]
	\centering
	\includegraphics[width=\textwidth]{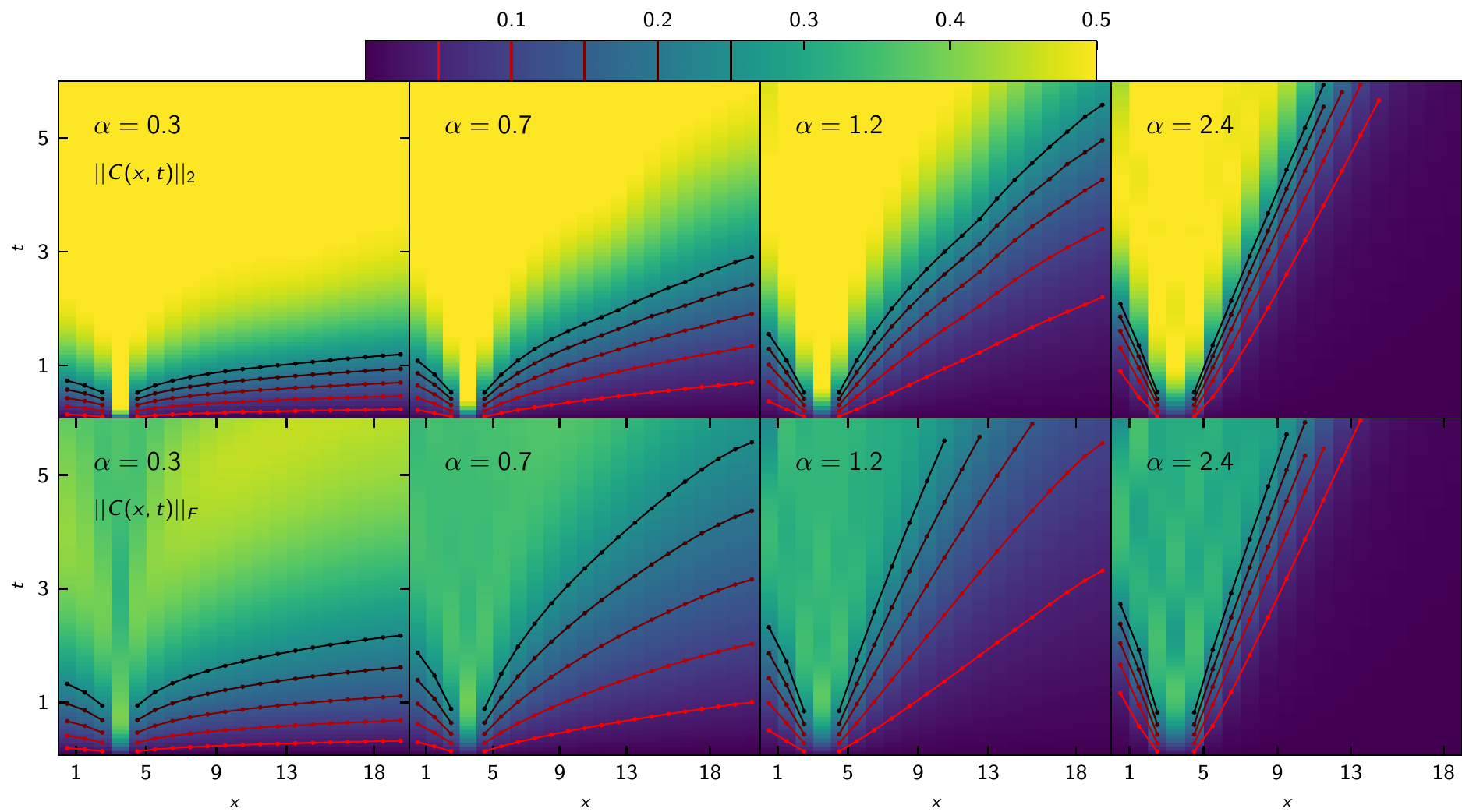}
	\caption{Norms of the commutator $\hat{C}(x,t)=[\hat S_i^z(t),S_{x}^z ]$ as a function of time $t$ and position $x$ in the chain for different interaction exponents $\alpha$. The first operator is located at $i=3$. The top row corresponds to the operator norm (largest singular value of $C(x,t)$), the bottom row shows the normalized Frobenius norm (root of mean of squared singular values of $C(x,t)$).
		The curves correspond to contour lines for different thresholds $\theta$, i.e. the solution $t_\theta(x)$ of the equation $\|C(x,t)\|=\theta$ with  $\theta=0.05, 0.1, 0.15, 0.2, 0.25$ for both norms. The interaction exponents $\alpha=0.3, 0.7, 1.2, 2.4$ correspond to system sizes $L=21, 22, 21, 20$ respectively. }
	\label{fig:color_map}
\end{figure*}
In the following, we analyze in detail the space-time profile of the operator norm of the commutator $\|\hat C(r,t)\|_2$ of the long range XXX chain \eqref{eq:model} and compare it to the case of the normalized Frobenius norm (OTOC), for which a very detailed analysis can be found in Ref. \onlinecite{luitz_emergent_2019}.
We provide additional complementary data for the XYZ chain in Appendix \ref{sec:xyz}.

\subsection{Causal space time region}

We start our analysis by providing a qualitative comparison of the two norms of the commutators $\hat{C}(r,t) = [\hat S_3^z(t), \hat S_{3+r}^z]$ for different range of the interaction $\alpha$ and all distances $r$ as a function of time. The synopsis of these results is shown in Fig. \ref{fig:color_map}, where the top row shows the operator norm and the bottom row the normalized Frobenius norm (OTOC), while columns correspond to different ranges of the interaction $\alpha=0.3, 0.7, 1.2, 2.4$. Both norms are shown on the same color scale. Full lines show contour lines of the space time profile for various thresholds $\theta$, extracted from the solution of the equation $C(r,t) = \theta$ to obtain the light ``cone'' $t_\theta(r)$. 
It is clear already from a visual inspection of the two norms that the essential behavior is identical. Both norms reveal a clear causal space time region outside of which the commutator is very small, which means that almost no quantum information is communicated at short times and long distances for all $\alpha$.

The contours are calculated for the same set of threshold values $\theta$ (indicated as vertical lines in the colorbar for clarity), clearly showing that the operator norm reaches a fixed threshold earlier than the Frobenius norm due to the property given by Eq.\eqref{eq:norm_ineq}. Since no signal can travel faster than governed by the operator norm, it strictly limits the amount of quantum communication outside the causal region. The comparison between the operator and the Frobenius norm shows that typical modes in the system [singular values of $\hat C(r,t)$] travel significantly slower than the fastest mode (maximal singular value), an effect which is particularly pronounced at small $\alpha$ as can be seen from a comparison of the contour lines between the two norms.

The overall shape of the contour lines appears to be identical (with different prefactors). For large $\alpha$, both norms are consistent with asymptotically linear light cones.

For intermediate $\alpha = 1.2$, and large thresholds (black contour lines), we observe a ``bump'' in the case of the operator norm, which is likely nonuniversal and stems from the reflection at the left edge of the system, therefore we focus on smaller thresholds in these cases, where reflection does not (yet) interfere due to the observed causality.

\subsection{Early time growth}

\begin{figure}[h]
	\centering
	\includegraphics{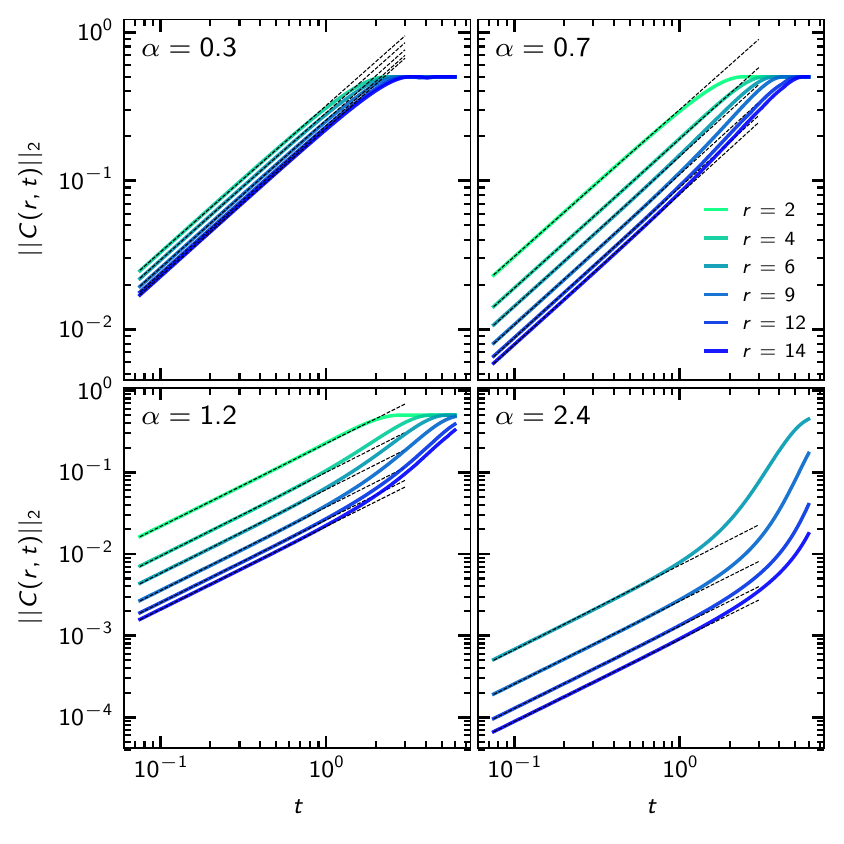}
	\caption{$C_2(r,t)$ as function time at fixed distance $r$ from site $i=3$. Dashed lines are linear power law fits to times $t<1$. The fitted lines are extended up to $t=3$. The system size is $L=22$.  }
	\label{fig:time_evolution}
\end{figure}

In Fig. \ref{fig:time_evolution} we analyze the growth of the operator norm $C_2(r,t)=\|\hat C(r,t)\|_2$ as a function of time $t$ for fixed distances $r$. The results are shown for a system of size $L=22$ and reveal very clearly that the operator norm grows linearly in time (linear growth shown by dashed black lines for comparison). We have checked that this short time behavior is converged in system size.
This is identical to the behavior of the normalized Frobenius norm (OTOC)\cite{luitz_emergent_2019} and
is expected in the short time perturbative regime when $t\ll1$. 
We show how this linear growth arises from short time perturbation theory in Sec. \ref{sec:PT}, where it becomes also clear that nearest neighbor interactions lead to a different (power law) short time behavior.

\subsection{Perturbation theory in the short time limit}\label{sec:PT}

At short times, we can use the Baker-Campbell-Hausdorff (BCH) formula
\begin{eqnarray}
\label{eq:bch formula}
e^{\hat{X}}\hat{Y}e^{-\hat{X}} = \sum_{m=0}^{\infty} \dfrac{1}{m!}[\hat{X},\hat{Y}]_m,
\end{eqnarray}
with $[\hat{X},\hat{Y}]_m = [\hat{X},[\hat{X},\hat{Y}]_{m-1}]$ and $[\hat{X},\hat{Y}]_0 = \hat{Y}$. Replacing $\hat{Y}=\hat{S}^z_{i}$ and $\hat{X} = \I t\hat{H}$ we get a perturbative expansion for time dependent Heisenberg operators: 
\begin{eqnarray}
\label{eq:st_bch}
\hat{S}^{z}_i(t) = \sum_{m=0}^{\infty} \dfrac{(\I t)^{m}}{m!}[\hat{H},\hat{S}^z_{i}]_m.
\end{eqnarray}
The commutator Eq. (\ref{eq:commutator_sz}) can then be written as
\begin{eqnarray}
\label{eq:commutator_bch}
[\hat{S}^{z}_i(t),\hat{S}^{z}_{i+r}] = \sum_{m=0}^{\infty} \dfrac{(\I t)^{m}}{m!}[[\hat{H},\hat{S}^z_{i}]_m,\hat{S}^{z}_{i+r}].
\end{eqnarray}

For systems with long range interactions, the commutator to linear order $[[\hat H, \hat S_i^z],\hat S_{i+r}^z]$ is nonzero for any distance $r$, and is therefore the leading order at short times, leading to a dominant term linear in $t$. 
For the long range XXX model Eq. \eqref{eq:model}, we obtain for $r>0$:
\begin{equation}
\label{eq:long_range_bch}
[\hat{S}^{z}_i(t),\hat{S}^{z}_{i+r}] =  \dfrac{\I t}{r^{\alpha}}\left(\hat{S}^{x}_i\hat{S}^{x}_{i+r}+\hat{S}^{y}_i\hat{S}^{y}_{i+r}\right)+\mathcal{O}(t^2).
\end{equation}
Therefore, the operator norm $C_2(r,t)$ to leading order in $t$ reads
\begin{equation}
\label{eq:op_norm_bch}
\|[\hat{S}^{z}_i(t),\hat{S}^{z}_{i+r}]\|_2 = \dfrac{t}{2r^{\alpha}}+\mathcal{O}(t^2).
\end{equation}
For any finite $\alpha$ the operator norm Eq. (\ref{eq:op_norm}) grows linearly in time and scales as $r^{-\alpha}$ at long distance and short times. We note that this perturbative behavior is true for any choice of the norm.
In Fig. \ref{fig:short_and_long_range} the exact time evolution (colored lines) is compared to the leading order Eq. \eqref{eq:op_norm_bch} (grey straight lines), yielding excellent agreement at short times.

On the other hand, Eq. \eqref{eq:commutator_bch} yields a very different behavior when interactions are limited to only nearest neighbors. In the short range limit $\alpha\rightarrow\infty$ the support of the nested commutator $[\hat{H},\hat{S}^z_{i}]_m$ grows by one lattice site at each $m$ term which makes $[[\hat{H},\hat{S}^z_{i}]_m,\hat{S}^{z}_{i+r}]$ vanishes for $m<r$. This can be seen more clearly by looking at the first term in the expansion $[\hat{H},\hat{S}^z_i]_1=i(\hat{S}^x_{i}\hat{S}^{y}_{i+1}-\hat{S}^y_{i}\hat{S}^{x}_{i+1})+i(\hat{S}^x_{i-1}\hat{S}^{y}_{i}-\hat{S}^y_{i-1}\hat{S}^{x}_{i})$ which has support only on sites $i-1,i,i+1$ and therefore $[[\hat{H},\hat{S}^z_{i}]_1,\hat{S}^{z}_{i+r}]$ vanishes as long as $r>1$. Higher order terms in the BCH formula for $r>1$ become only nonzero if a string of nontrivial Pauli matrices of length $r$ is generated between sites $i$ and $i+r$, and thus
the leading order in the BCH formula reads for nearest neighbor interactions
\begin{eqnarray}
\label{eq:short_range_bch}
[\hat{S}^{z}_i(t),\hat{S}^{z}_{i+r}] \propto  \dfrac{t^{r}}{r!}\hat{\mathcal{O}}(1),
\end{eqnarray}
where $\hat{\mathcal{O}}(1)$ is given by the operator norm of a sum of Pauli strings of length $r$. In other words, at short times and outside the light cone the operator norm grows as a power law in time with an exponent given by the distance between the two operators. This is a quite general result, valid for any pair of local operators that are separated by a distance $r$ larger than the support of the most extended term in the Hamiltonian. In Fig. \ref{fig:short_range_time} the exact time evolution of Eq. \eqref{eq:op_norm}  for the XXX short range model is compared to the perturbation theory result Eq. \eqref{eq:short_range_bch}. The power law growth $C_2(r,t)\propto t^r$ is in excellent agreement with the exact calculation at short times.

\subsection{Role of the nearest neighbor part of the Hamiltonian at large $\alpha$}

\begin{figure}[h]
	\centering
	\includegraphics{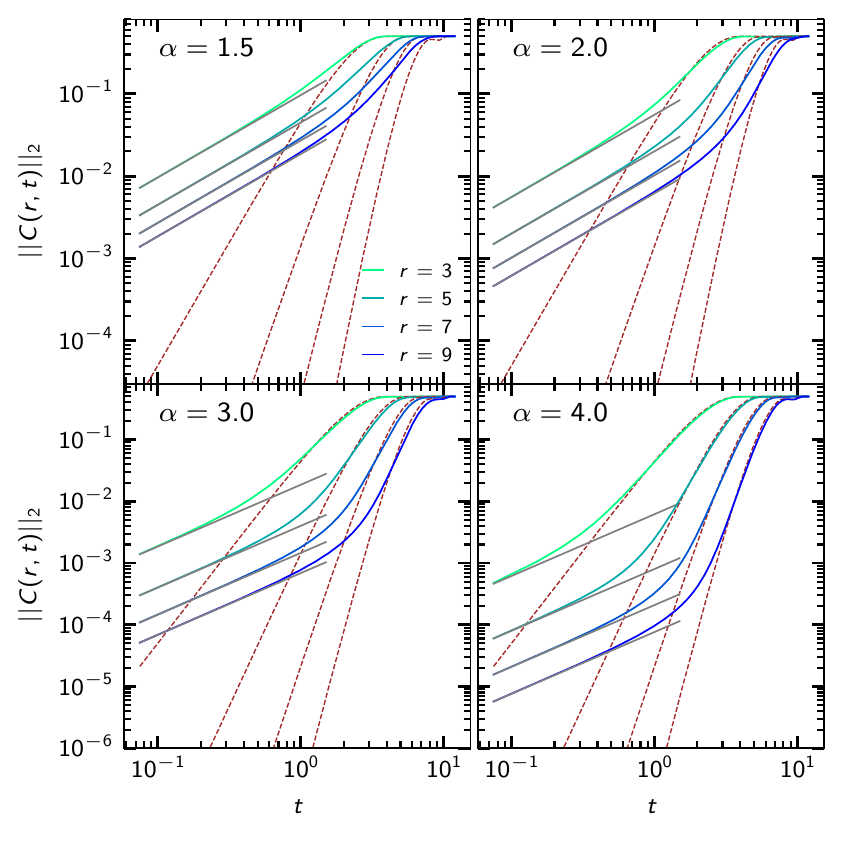}
	\caption{Time evolution of $C_2(r,t)$ at fixed distance $r=3,5,7,9$ from $i=3$ and $\alpha=1.5,2.0,3.0,4.0$ for nearest neighbor (dashed lines) and long range interactions. Grey lines are the leading order in perturbation theory given by Eq. \eqref{eq:op_norm_bch}. The system size is $L=14$. }
	\label{fig:short_and_long_range}
\end{figure}

\begin{figure}[h]
	\centering
	\includegraphics{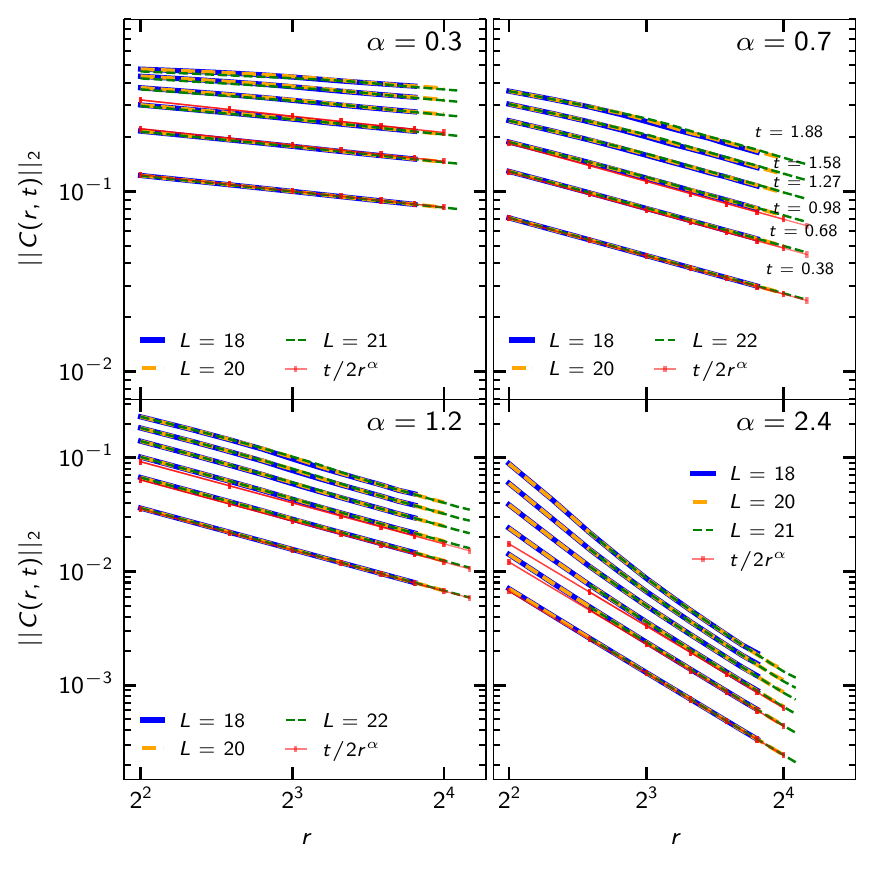}
	\caption{Commutator as function of distance $r$ from site $i=3$ at times $t=0.38,0.68,0.98,1.27,1.58,1.88$ (descending order in the plots). The overall behavior seems to be power law at any fixed time. The trends of the smaller sizes are followed off by the larger ones. }
	\label{fig:distance}
\end{figure}

For large values of $\alpha$, there is a significant speedup of the growth of the commutator norm and a clear departure from the linear growth of $C_2(r,t)$ at intermediate times (cf. Fig. \ref{fig:time_evolution} lower panels). On the other hand, when $\alpha$ is large enough the commutator $C_2(r,t)$ is expected to exhibit a similar behavior as a short range interacting system, which corresponds to the limit $\alpha\rightarrow\infty$ in Eq. (\ref{eq:model}). The Hamiltonian of the long range model contains the nearest neighbor part plus longer distance couplings, decaying as $1/r^\alpha$, which are strongly suppressed for $\alpha \gg 1$. Therefore, a dominant effect of the nearest neighbor part is expected for large $\alpha$ \cite{kloss_spin_2019,luitz_emergent_2019}.
In Fig. \ref{fig:short_and_long_range} we compare $C_2(r,t)$ for the long range model (full colored lines) to the nearest neighbor model (red dashed) at fixed distances $r$. At short times, the long range model shows the perturbative growth $r^{-\alpha} t/2$ for all $\alpha$, and speeds up at intermediate times. The initial growth is significantly faster than in the case of nearest neighbor interactions. For nearest neighbor interactions, the operator norm $C_2(r,t)$ grows much faster due to the large power law $\propto t^r$. Therefore, at later times and for $\alpha>1$, the nearest neighbor part catches up and dominates the overall growth of the commutator and leads to an asymptotic linear light cone.

Focusing only on large $\alpha$, the operator norm of the commutator exhibits two kinds of grows: linear at short times (see Fig. \ref{fig:time_evolution}) and short-ranged like at intermediate times (see Fig. \ref{fig:short_and_long_range}). The short-range time evolution of $C_2(r,t)$ is well characterized by $(t/r)^r$\cite{wang_tightening_2020}, while the long-range part is described by Eq \eqref{eq:op_norm_bch} $t/r^{\alpha}$ in the limit $t\ll1$. These two results can be combined into a single expression: 
\begin{eqnarray}
\label{eq:asymptotic_time}
C_2(r,t) \propto \dfrac{t}{2r^{\alpha}} + \mathcal{O}(1) \left(\dfrac{t}{r}\right)^r.
\end{eqnarray}
At short times the linear term on the right hand side is always dominant, at intermediate time the second term become dominant and the dynamics is short-range like. 

At long distances there is a clear transition from a linear light cone for $\alpha>1$ to a power law light cone at $\alpha<1$, which can be understood with the following reasoning. The asymptotic form of $C_2(r,t)$ in Eq. \eqref{eq:asymptotic_time} grows monotonically and the two terms compete. The light cone is given by the set of times $t_\theta(r)$ as a function of distance $r$, for which $C_2(r,t)$ reaches a threshold value $\theta$, i.e. $C_2(r,t_\theta(r))=\theta$. It is clear that $t_\theta(r)\leq t_c(r)= 2 \theta r^\alpha$, since this is the time the first (linear in $t$) term needs to reach the threshold. If the second (power law in $t$) term reaches the threshold first, we get a linear light cone, otherwise we get a power law light cone. We can estimate the power law term at long distances and $t\leq t_c$ by $t^r/r^r\leq t_c^r/r^r=(2\theta r^\alpha)^r/r^r$. Therefore, at $t_c$, this term diverges for $\alpha>1$ and $r\to\infty$, and overwhelms the linear term, leading to a linear light cone $t_\theta(r)\propto r$. For $\alpha<1$, and $r\to\infty$, this term is irrelevant and we are left with a power law light cone $t_\theta(r)\propto r^\alpha$. Analogously, the linear term is bounded by the time when the short range front reaches the threshold, i.e.  $t/2r^{\alpha}\le r\theta^{1/r}/2r^{\alpha}$. The bounding term vanishes in the limit $r\rightarrow\infty$ and $\alpha>1$ and diverges otherwise, which agrees with the bound to the short range term. This behavior is consistent with the numerical observation in Fig. \ref{fig:color_map}.

\subsection{Long distance decay}

\begin{figure}[h]
	\centering
	\includegraphics{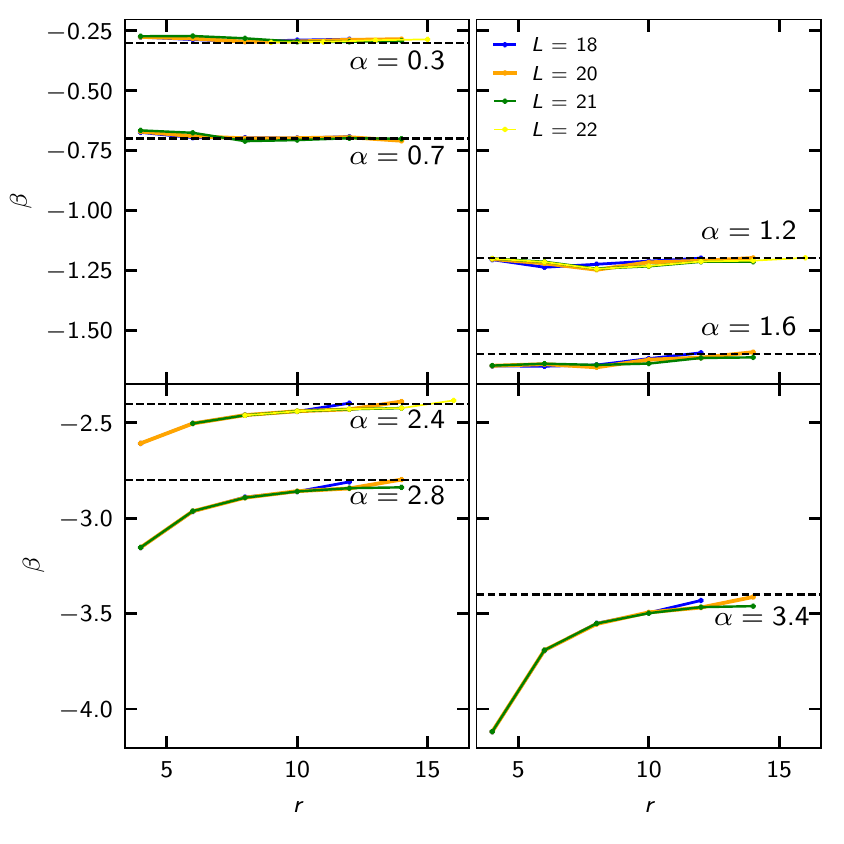}
	\caption{Tail exponent of long distance decay (see Fig. \ref{fig:distance}) computed from the discrete logarithmic derivative $\beta=\partial_{\log r} \log C_2(r,t)$ as a function of distance $r$ for different system sizes $L$ at fixed time $t=0.67$. The asymptotic exponent $\beta$ converges towards the interaction exponent $\alpha$ (dashed horizontal lines) for all values of $\alpha$ at long distances.  }
	\label{fig:beta}
\end{figure}

In Fig. \ref{fig:distance} we analyze the behavior of $C_2(r,t)$ at long distances $r$ outside the ``light cone". It falls off as a power law at long distances, with an exponent that asymptotically approaches the interaction exponent $\alpha$ (see quantitative analysis using a discrete logarithmic derivative in Fig. \ref{fig:beta}). 
The same behavior was found previously \cite{luitz_emergent_2019} for the normalized Frobenius norm. 
This power law decay $r^{-\alpha}$ is in perfect agreement with the prediction from perturbation theory in the short time limit given by Eq. \eqref{eq:op_norm_bch}, and seems to be valid asymptotically outside the causal region. 

This analysis confirms the validity of short time perturbation theory (which is valid for any matrix norm) and shows that the asymptotic shape of the tails (outside the causal region) of long range interacting spin chains is given by $C_2(r,t)\propto t/r^{\alpha}$.

\section{Conclusions}

The operator norm $C_2(r,t)$ of the commutator Eq. \ref{eq:conmutator_def} has been examined in the XXX chain with long range interactions falling off as a power law $r^{-\alpha}$ with distance $r$ and interaction exponent $\alpha$. In order to reach large enough systems to check the convergence of our results with system size, we introduce a Krylov space method for the direct calculation of the operator norm of the commutator $C(r,t)$.
We find a linear growth in $t$ at early time $t$ and a long distance decay outside the causal region given by $r^{-\alpha}$.
Both the normalized Frobenius norm (directly related to OTOCs) and operator norm which correspond to the average and fastest information spreading modes in the system have the same asymptotic behavior (with different prefactors) of $t/r^\alpha$ at long distance and short time, which is strikingly different from systems with nearest neighbor interactions, which instead exhibit a leading growth as a power law in time $\propto t^r$ at short times and long distances. 

For $\alpha>1$, the information front is dominated by the contribution from the nearest neighbor part of the Hamiltonian, which overtakes the initial linear growth of $C_2(r,t)$, inducing a linear light cone.
These results are confirmed using a slightly different XYZ model in appendix \ref{sec:xyz}.

We conclude that the findings from the study of out of time order correlators in Ref. \onlinecite{luitz_emergent_2019} provide information about Lieb Robinson bounds and agree with the behavior of the fastest spreading information mode in the system.

\section{Acknowlegments}
We acknowledge financial support from the Deutsche Forschungsgemeinschaft (DFG) through SFB  1143 (project-id  247310070).

\clearpage

\appendix

\section{Perturbative treatment at short times}
\label{appendix:perturbation}

\begin{figure}[h]
	\centering
	\includegraphics{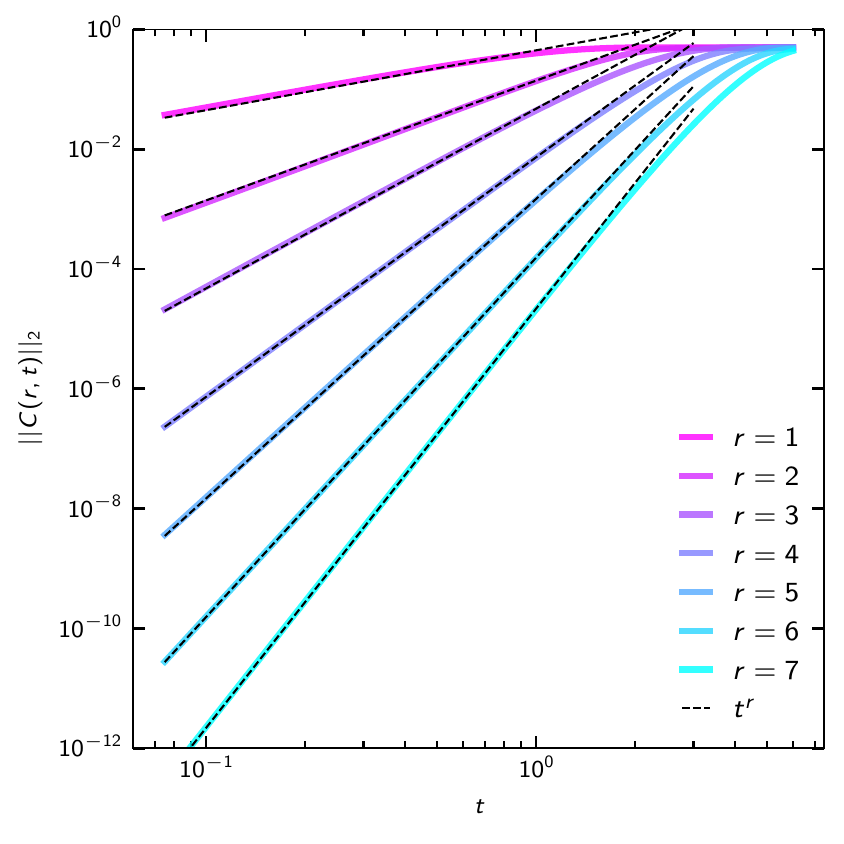}
	\caption{Time evolution of the operator norm $C_2(r,t)$ at fixed distance $r$ for the nearest neighbor XXX chain with system size $L=18$. Black dashed lines are power low fits $C_2(r,t)\sim t^r$ for the corresponding $r$.   This is the behavior obtained in Eq. (\ref{eq:short_range_bch}) from perturbation theory. }
	\label{fig:short_range_time}
\end{figure}

\begin{figure}[h]
	\centering
	\includegraphics{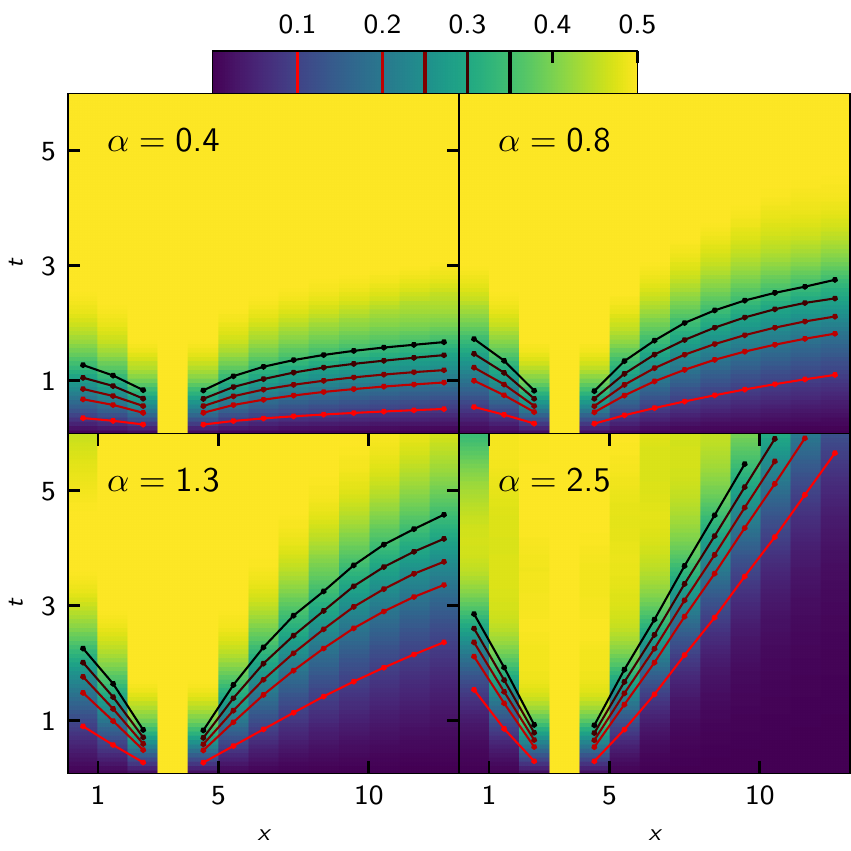}
	\caption{Spreading of operator norm $C_2(x,t)=||[S^z_{i}(t),S^x_{x}]||_2$ in the XYZ model \eqref{eq:model_xyz} with $i=3$, over time and space for $\alpha=0.4,0.8,1.3,2.5$. Continuous lines are contour lines given by the equation  $\|C(x,t)\|=\theta$ with $\theta=0.1,0.2,0.25,0.3,0.35$. System size is $L=14$. The shape of the "light cone" and their countour lines are very similar to the ones discussed in the main text (see Fig. \ref{fig:color_map})  }
	\label{fig:color_map_xyz}
\end{figure}

 In the main text, section \ref{sec:PT}, the Baker-Campbell-Hausdorff (BCH) formula was employed for treating the time evolution of Eq. \eqref{eq:conmutator_def} at short times. From this analysis, the leading order of the operator norm $C_2(r,t)\approx t^r/r!$ was obtained. In Fig. \ref{fig:short_range_time} the exact time evolution $C_2(r,t)$ is shown along with the leading order in the BCH formula (dashed lines), with excellent agreement at short times.

\section{Results for XYZ model}

\begin{figure}[h]
	\centering
	\includegraphics{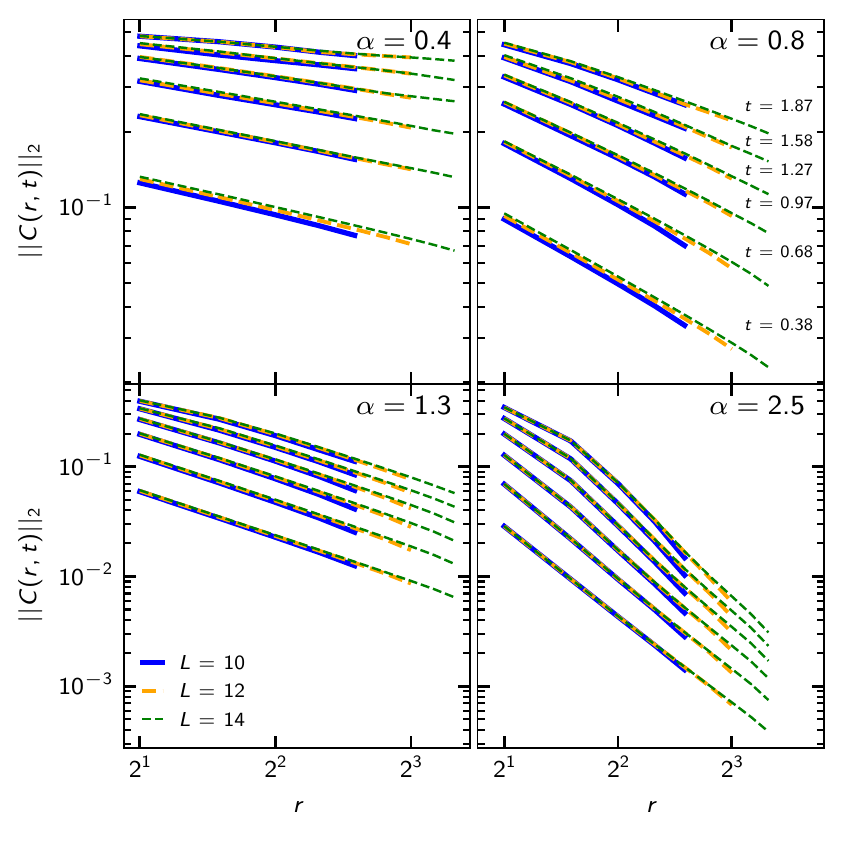}
	\caption{Spatial decay of $C_2(r,t)$ in the XYZ model \eqref{eq:model_xyz} at fixed time $t_0=0.38,0.68,0.98,1.27,1.58,1.88$. Similarly to XXX model treated in the main text, there is a power law decay (see Fig. \ref{fig:distance}).}
	\label{fig:distance_xyz}
\end{figure}

\begin{figure}[h]
	\centering
	\includegraphics{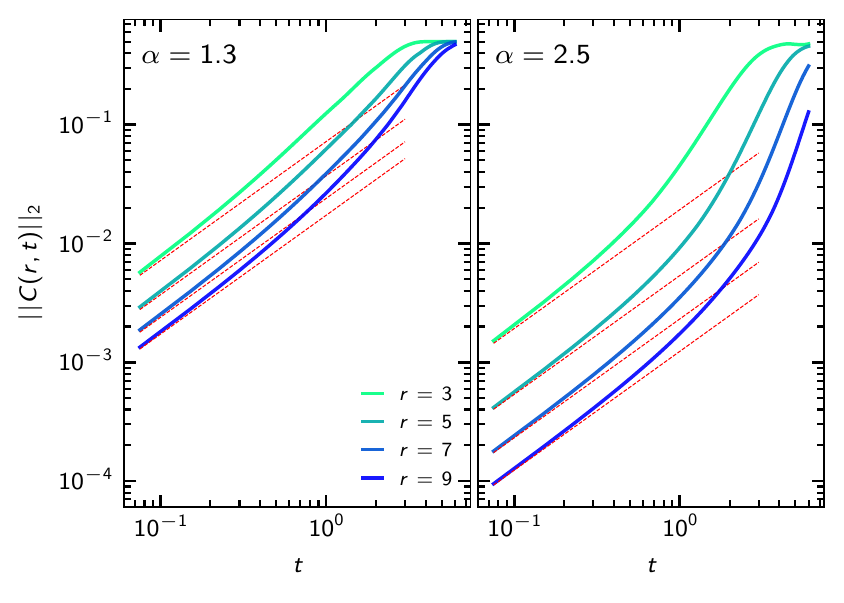}
	\caption{Time evolution of $C_2(r,t)$ in the XYZ model \eqref{eq:model_xyz} at fixed distance $r_0$ and system size $L=14$. Dashed lines are the asymptotic form $t/4r^{\alpha}$ at short times $t<1$. The overall behavior is again similar to what was found for the XXX model (see Fig. \ref{fig:time_evolution}) }
	\label{fig:time_xyz}
\end{figure} 
\label{sec:xyz}
In order to test the universality of the results presented in the main text, we have also performed similar calculations for the long range XYZ Heisenberg model: 
\begin{equation}
\label{eq:model_xyz}
H = \sum_{i<j}\!\dfrac{1}{|i-j|^{\alpha}}\!\left(J_x\hat{S}^{x}_{i}\hat{S}^{x}_{j}\!+J_y\hat{S}^{y}_{i}\hat{S}^{y}_{j}+J_z\hat{S}^{z}_{i}\hat{S}^{z}_{j}\right),
\end{equation}
with parameters $J_x=0.9,J_y=1.2,J_z=0.7$. The XXX model study in the main text is recovered by setting $J_x=J_y=J_z=1$. The XYZ model does not have $U(1)$ symmetry, therefore the Hamiltonian is not block diagonal and we must deal with the full Hilbert space dimension $\mathcal{N}=2^L$. Analogously to the main text, we study the operator norm of the commutator
\begin{eqnarray}
\label{eq:conmutator_def_xyz}
C_2(r,t) = ||[\hat{S}^z_i(t),\hat{S}^x_{i+r}]||_2.
\end{eqnarray}
The difference compared to Eq. \eqref{eq:op_norm} lies in the static operator that is now $\hat{S}^x_{i+r}$. The time evolution and operator norm computation are carried out using exact diagonalization. 

We analyze the behavior of Eq. \eqref{eq:conmutator_def_xyz} as function of both $r$ and $t$. In Fig. \ref{fig:color_map_xyz} the causal regions of $C_2(r,t)$ are shown. For all values of $\alpha$ the overall shape of the causal region is the same as for the XXX long range model (see Fig. \ref{fig:color_map}). Small $\alpha$ exhibit fast spreading with power law causal regions and large $\alpha$ approach the linear light cone limit $\alpha\rightarrow\infty$. The long distance decay is also similar to what has been found in the main text (see Fig. \ref{fig:distance_xyz} and \ref{fig:distance}) outside the light ``cone" there is a power law decay of $C_2(r,t)$. Time evolution at fixed distance is displayed in Fig. \ref{fig:time_xyz} and is compatible with linear growth at short times $t<1$ which was found also in the XXX version (see Fig. \ref{fig:time_evolution}). In conclusion, the main features of both operator norm discussed in the main text are the same when considering a different commutator $[\hat{S}^z_i(t),\hat{S}^x_{i+r}]$ and a different long range model, namely the long range XYZ model. As expected only the interaction exponent $\alpha$ is crucial for characterizing the dynamics of the commutator Eq. \eqref{eq:conmutator_def}

Applying perturbation theory Eq. \eqref{eq:commutator_bch} up to first order we get the following expression for the commutator when $r>0$:
\begin{eqnarray}
\label{eq:commitator_xyz_bch}
[\hat{S}^z_i(t),\hat{S}^x_{i+r}]=\dfrac{t}{r^\alpha}S^{z}_{i+r}S^{x}_{i}+\mathcal{O}(t^2),
\end{eqnarray}
yielding the following asymptotic form for the operator norm:
\begin{eqnarray}
\label{eq:xyz_bch}
 ||[\hat{S}^z_i(t),\hat{S}^x_{i+r}]||_2=\dfrac{t}{4r^\alpha}+\mathcal{O}(t^2).
\end{eqnarray}
In Fig. \ref{fig:time_xyz} this asymptotic form is compared with the exact time evolution yielding very good agreement. In conclusion, the BCH formula also predicts the short time behavior for the XYZ model.

\bibliography{op-norm,quantum_information}

\clearpage
\appendix

\end{document}